\documentclass[aps,pra,preprint,superscriptaddress,longbibliography]{revtex4-2}
\usepackage[T1]{fontenc}
\usepackage[utf8]{inputenc}
\usepackage{amsmath,amssymb,bm}
\usepackage{graphicx}
\usepackage{hyperref}
\usepackage{etoolbox}

\makeatletter
\patchcmd{\frontmatter@abstract@produce}
    {\preprintsty@sw}{\@secondoftwo}{}
    {\PackageError{arXivV3}{Could not adjust abstract placement}{Check the REVTeX version.}}
\makeatother
\hypersetup{colorlinks=true,urlcolor=blue,linkcolor=black,citecolor=black}
\begin{document}
\title{Grid-state deformation in a no-jump non-Hermitian bosonic dimer}
\author{B.~M. Rodriguez-Lara}
\email{blas.rodriguez@gmail.com}
\affiliation{Universidad Polit\'ecnica Metropolitana de Hidalgo, Tolcayuca, Hidalgo 43860, Mexico}
\author{H. Ghaemi-Dizicheh}
\affiliation{Department of Physics and Astronomy, University of Texas Rio Grande Valley, Edinburg, TX 78539, United States of America}
\author{S. Dehdashti}
\affiliation{Emmy-Noether Group Theoretical Quantum Systems Design, Technical University of Munich, Munich, Germany}
\author{A. Hanke}
\email{andreas.hanke@utrgv.edu}
\affiliation{Department of Physics and Astronomy, University of Texas Rio Grande Valley, Edinburg, TX 78539, United States of America}
\author{A. Touhami}
\affiliation{Department of Physics and Astronomy, University of Texas Rio Grande Valley, Edinburg, TX 78539, United States of America}
\author{J. N\"otzel}
\affiliation{Emmy-Noether Group Theoretical Quantum Systems Design, Technical University of Munich, Munich, Germany}

\begingroup
\linespread{1}\selectfont
\begin{abstract}
    We study the no-jump evolution of ideal grid states in a lossy bosonic dimer with differential decay.
    The effective non-Hermitian quadratic dynamics induces a complex symplectic flow in phase space that deforms both the primitive lattice vectors and the origin seed.
    The average decay rate controls common-mode attenuation, while coherent hopping and differential decay control the reduced dimer deformation.
    For resonant modes, the reduced displacement flow is oscillatory in the elliptic regime, linear at the exceptional point, and exponentially separating in the hyperbolic regime.
    Although projected lattice areas can change, the deformation comes from a determinant-one complex symplectic flow on the
    complexified phase space.
    For a Gaussian regularization of the origin seed in the resonant dimer,
    we derive the associated complex width matrix and
    show that the seed remains square-integrable and Gaussian
    for every positive initial width, all non-negative times, and all physical
    loss rates.
    For an initial two-mode qunaught product state, the lossless limit recovers the standard beam-splitter generation of a square Gottesman--Kitaev--Preskill Bell pair, while the no-jump dynamics produces its non-Hermitian deformation.
    The full no-jump probability for a normalized finite-energy input sets the postselection cost and depends on its excitation-number distribution and the reduced dimer dynamics.
\end{abstract}
\maketitle
\endgroup
\clearpage
\section{Introduction}
\label{sec:Sec1}

Bosonic continuous-variable systems provide a setting for quantum information processing, where Gaussian states, Gaussian operations, and bosonic channels supply the basic language for encoding, control, and noise analysis \cite{Braunstein2005p513,Weedbrook2012p621}.
Within this setting, Gottesman--Kitaev--Preskill (GKP) states encode finite-dimensional logical information in the phase space of bosonic modes \cite{Gottesman2001p016316,Glancy2006p012325}.
A GKP state forms a coherent superposition of displaced copies of an elementary seed state over a phase-space lattice.
The lattice fixes the displacement pattern of the grid, while the seed fixes the elementary state repeated over it \cite{Conrad2022p648,Royer2022p010335}.
Ideal GKP states use singular, infinitely squeezed peaks, whereas physical approximations replace them with finitely squeezed Gaussian wavepackets and a decaying envelope over the lattice.
Lattice translational symmetry lets the code detect and correct small displacement errors, which protects the encoded information against bosonic noise \cite{Gottesman2001p016316,Glancy2006p012325}.
The role of grid states in bosonic error correction \cite{Grimsmo2021p020101,Brady2024p100496}, lattice-based code constructions \cite{Conrad2022p648,Royer2022p010335}, and measurement-based or concatenated fault-tolerant architectures \cite{Larsen2021p030325,Walshe2021p062427,Noh2022p010315} makes the generation and dynamical understanding of grid encodings a central problem in continuous-variable quantum information.
Existing work has developed phase-space and lattice-based descriptions of GKP states and related bosonic encodings, together with analyses of their generation, error correction, and response to loss or damping \cite{Mensen2021p022408,Walshe2020p062411,Hastrup2023p052413,Harris2025p042417,Zheng2023p012603}.
More broadly, bosonic modes also enable hybrid
continuous--discrete-variable encodings and autonomous quantum-error-correction
schemes based on engineered dissipation \cite{Cho2026}.
Dissipative dynamics can deform GKP lattices \cite{Wang2025p021003},
and two-qubit operations designed for ideal GKP codes can
generate undesired entanglement in finite-energy implementations
\cite{Rojkov2024p100601}.

Open bosonic systems also admit a trajectory description, where the continuous evolution between quantum jumps is generated by an effective non-Hermitian Hamiltonian \cite{Dalibard1992p580,Carmichael1993p2273,Plenio1998p101}.
For monitored dynamics, the no-jump branch is the conditioned evolution associated with the absence of detected decay events.
Quantum error correction for monitored dynamics must account
for the conditional backaction associated with both detected jumps and
no-jump measurement records
\cite{Ahn2003p052310,Michael2016p031006}.
Non-Hermitian and exceptional-point physics has clarified how gain, loss, and mode coupling organize the spectral and dynamical structure of open systems, particularly in optical and photonic settings \cite{Bender1998p5243,Ashida2020p015005,ElGanainy2018p11,Miri2019peaar7709}.

Beyond their role as minimal models of non-Hermitian spectral
transitions, dimer structures also provide platforms for functional wave and
energy control. Non-Hermitian topological dimer chains combine
exceptional-point physics with topological edge modes to enable functional
applications, including enhanced sensing \cite{Guo2021p574}
and robust wireless power transfer \cite{Song2021p014009}.
These studies illustrate how engineered coupling,
gain and loss, and topology can provide functional control in dimer systems.
The present work considers a complementary quantum-information setting, in
which conditioned non-Hermitian dimer dynamics controls the deformation of
bosonic grid encodings.

However, to our knowledge, no exact encoding-level description evolves the GKP lattice and its associated seed as a single phase-space object under conditioned non-Hermitian dynamics.
Non-Hermitian Gaussian dynamics provides exact propagation laws for Gaussian centers, widths, and covariance data under quadratic effective Hamiltonians, but it does not give an evolution law for the primitive generators of a non-Gaussian phase-space comb \cite{Graefe2011p060101,Christie2022p455302}.
Conversely, phase-space and Zak-space approaches to GKP codes give powerful lattice descriptions under Gaussian operations, but they mostly address unitary Gaussian transformations or noise-channel descriptions rather than post-selected complex-symplectic no-jump flows \cite{Mensen2021p022408,Pantaleoni2023p062611}.
Similarly, lossy quadratic bosonic dimers and coupled-oscillator master equations display effective non-Hermitian dynamics and exceptional-point structure, yet this spectral regime structure has not been translated into code-level geometric deformation of grid lattices and their regularized seeds \cite{Teuber2020p042124,Sanchez2023p5435,Minganti2019p062131}.

To address this gap, we use a lossy bosonic dimer with coherent hopping and differential decay as a model system.
This two-mode setting connects directly with beam-splitter interactions and coupled-mode realizations in integrated photonics \cite{Politi2008p646,QuirozJuarez2019p862,Saxena2023p5260}, ultracold atoms \cite{Albiez2005p010402}, trapped ions \cite{Gan2020p170502}, optomechanics \cite{Weaver2017p824,JaramilloAvila2020p1761}, and superconducting bosonic hardware \cite{Chapman2023p020355,Lu2023p5767}.
For the no-jump branch of the lossy bosonic dimer, we derive the exact conditioned evolution of the primitive lattice vectors and the origin-seed width matrix of the GKP lattice.
This construction connects no-jump non-Hermitian dimer physics with the deformation of multimode grid-state encodings.
Lossy beam-splitter dynamics can also realize exceptional-point structure under postselection \cite{QuirozJuarez2019p862,Ghaemidizicheh2025p26329,RodriguezLara2026p1604}, which provides a non-Hermitian precedent for the dimer dynamics considered here.

The conditioned deformation of the encoding follows from the dimer parameters alone.
For resonant modes, the reduced displacement flow is
oscillatory in the elliptic regime, linear at the exceptional point, and
exponentially separating in the hyperbolic regime.
The average decay rate
fixes the common-mode attenuation scale, whereas the full no-jump
probability also depends on the input state's excitation-number
distribution and, for $\gamma_{-}\neq 0$, on the differential-loss
dynamics through the reduced propagator. This full probability, rather
than the common-mode attenuation scale alone, sets the postselection
cost.
Although dissipative dynamics can deform GKP lattices \cite{Wang2025p021003}, our classification gives an exact encoding-level account of the conditioned lattice-and-seed deformation through the spectrum of the reduced dimer.
This structure enables lattice-geometry engineering through the hopping strength, the relative hopping phase, and the decay rates of the modes.

The rest of the paper is organized as follows.
In Sec.~\ref{sec:Sec2} we introduce the lossy bosonic dimer and derive its no-jump branch from an effective non-Hermitian Hamiltonian.
In Sec.~\ref{sec:Sec3} we separate ideal grid-state evolution into lattice evolution and seed evolution.
In Sec.~\ref{sec:Sec4} we analyze the reduced dimer flow and its exceptional-point boundary.
In Sec.~\ref{sec:Sec5} we study the seed dynamics and show that a complex width matrix governs its regularized evolution.
In Sec.~\ref{sec:Sec6} we evolve a two-mode qunaught product state, recover the standard beam-splitter generation of a square GKP$+$ Bell pair in the lossless limit, and obtain its non-Hermitian deformation under no-jump dynamics.
In Sec.~\ref{sec:experiments} we discuss
candidate experimental
platforms and representative implementation parameters.
We discuss our results and conclude in Sec.~\ref{sec:discussion}.

\section{No-jump lossy bosonic dimer}
\label{sec:Sec2}

We consider a bosonic dimer formed by two degenerate modes with coherent hopping,
\begin{align}
    \label{eq:HBD}
    \hat{H}_{\mathrm{BD}} =&~ \hbar \omega_{0} \hat{n}_{+} + \hbar g \left( e^{i\phi} \hat{a}_{1} \hat{a}_{2}^{\dagger} + e^{-i\phi} \hat{a}_{1}^{\dagger} \hat{a}_{2} \right),
\end{align}
where $\hat{a}_{j}$ and $\hat{a}_{j}^{\dagger}$ annihilate and create excitations in mode $j = 1, 2$, $\hat{n}_{j} = \hat{a}_{j}^{\dagger}\hat{a}_{j}$, and $\hat{n}_{\pm} = \hat{n}_{1} \pm \hat{n}_{2}$ give the total and differential excitation numbers.
The parameter $\omega_{0}$ is the common mode frequency, $g > 0$ is the hopping strength, and $\phi \in \left[ 0, 2\pi \right)$ is the relative hopping phase.
Each mode couples to an independent zero-temperature environment that produces single-excitation loss with rate $\gamma_{j} \ge 0$.
The density operator satisfies the Lindblad master equation \cite{Lindblad1976p119,Gorini1976p821},
\begin{align}
    \label{eq:Lindblad}
    \frac{d}{dt} \hat{\rho}(t) =&~ -\frac{i}{\hbar} \left[ \hat{H}_{\mathrm{BD}}, \hat{\rho}(t) \right] + \sum_{j=1}^{2} \gamma_{j} \left( \hat{a}_{j} \hat{\rho}(t) \hat{a}_{j}^{\dagger} - \frac{1}{2} \left\{ \hat{n}_{j}, \hat{\rho}(t) \right\} \right),
\end{align}
where the square and curly brackets denote the commutator and anticommutator, and the annihilation operators act as jump operators for single-excitation loss.
Collecting the anticommutator terms with the Hamiltonian rewrites the master equation as
\begin{align}
    \label{eq:LindbladEff}
    \frac{d}{dt} \hat{\rho}(t) =&~ -\frac{i}{\hbar} \left[ \hat{H}_{\mathrm{eff}} \hat{\rho}(t) - \hat{\rho}(t) \hat{H}_{\mathrm{eff}}^{\dagger} \right] + \sum_{j=1}^{2} \gamma_{j} \hat{a}_{j} \hat{\rho}(t) \hat{a}_{j}^{\dagger},
\end{align}
with the effective non-Hermitian Hamiltonian
\begin{align}
    \label{eq:Heff}
    \frac{\hat{H}_{\mathrm{eff}}}{\hbar} =&~ \frac{\hat{H}_{\mathrm{BD}}}{\hbar} - \frac{i}{2}\gamma_{+}\hat{n}_{+} - \frac{i}{2}\gamma_{-}\hat{n}_{-},
\end{align}
where $\gamma_{\pm} = \left( \gamma_{1} \pm \gamma_{2} \right)/2$ are the average and differential decay rates, with $\gamma_{+} \ge \lvert \gamma_{-} \rvert$.
This form separates common-mode attenuation from the relative loss imbalance between the two modes.
Our lossy bosonic dimer is the dissipative Bose--Hubbard dimer in the absence of on-site interactions; see Ref.~\cite{Kordas2015p2127} for a review of dissipative Bose--Hubbard systems and Refs.~\cite{Giraldo2020p043009,Giraldo2022p385,Secli2021p063056,Muraev2023p117} for driven-dissipative dimer dynamics.

In the quantum-trajectory picture, the effective non-Hermitian Hamiltonian generates the deterministic evolution between quantum jumps \cite{Dalibard1992p580,Carmichael1993p2273,Plenio1998p101}.
This conditional evolution decreases the state norm.
The recycling term, $\sum_{j} \gamma_{j} \hat{a}_{j} \hat{\rho}(t) \hat{a}_{j}^{\dagger}$, transfers the norm lost from the conditioned branch from the $n_{+}$-excitation sector to the $\left( n_{+} - 1 \right)$-excitation sector, so the ensemble-averaged density operator keeps unit trace.
Reservoir engineering uses this structure to shape jump operators and drive open systems into target pure or entangled steady states \cite{Poyatos1996p4728,Verstraete2009p633}.

Conditioning on trajectories with no jumps removes the recycling term from the Lindblad equation.
For an initial pure state, the conditioned state remains pure,
\begin{align}
    \tilde{\rho}(t) =&~ \lvert \tilde{\psi}(t) \rangle \langle \tilde{\psi}(t) \rvert,
\end{align}
where tildes denote unnormalized conditioned quantities.
The unnormalized conditioned state vector satisfies the non-Hermitian Schr\"odinger equation
\begin{align}
    i\hbar \frac{d}{dt} \lvert \tilde{\psi}(t)\rangle =&~ \hat{H}_{\mathrm{eff}} \lvert \tilde{\psi}(t)\rangle.
\end{align}
The non-Hermitian evolution changes the norm of the conditional state,
\begin{align}
    P_{0}(t) =&~ \langle \tilde{\psi}(t) \vert \tilde{\psi}(t) \rangle,
\end{align}
which decreases monotonically according to
\begin{align}
    \frac{d}{dt} P_{0}(t) =&~ -\sum_{j=1}^{2}\gamma_{j} \langle \tilde{\psi}(t) \vert \hat{n}_{j} \lvert \tilde{\psi}(t)\rangle.
\end{align}
For a normalized initial state, $P_0(t)$ is the
probability that no jump has occurred up to time $t$ and thus
determines the postselection cost of the conditioned dynamics.

The effective Hamiltonian commutes with the total excitation number, $\left[ \hat{H}_{\mathrm{eff}}, \hat{n}_{+} \right] = 0$, because the free term, the hopping term, and the loss imbalance preserve $n_{+}$.
The evolution operator factorizes exactly,
\begin{align}
    \label{eq:Factorization}
    e^{-\frac{i}{\hbar} \hat{H}_{\mathrm{eff}} t} =&~ e^{-i \omega_{0} \hat{n}_{+} t} \, e^{- \frac{1}{2} \gamma_{+} \hat{n}_{+} t } \, e^{-\frac{i}{\hbar} \hat{H}_{n_{+}} t },
\end{align}
where the reduced non-Hermitian Hamiltonian contains the coherent hopping and the relative loss imbalance,
\begin{align}
    \label{eq:Hred}
    \frac{\hat{H}_{n_{+}}}{\hbar} =&~ g \left( e^{i\phi} \hat{a}_{1} \hat{a}_{2}^{\dagger} + e^{-i\phi} \hat{a}_{1}^{\dagger} \hat{a}_{2} \right) - \frac{i}{2} \gamma_{-} \hat{n}_{-}.
\end{align}
The first factor in Eq.~\eqref{eq:Factorization} produces a global phase-space rotation at the common mode frequency, which we remove by working in the frame rotating at $\omega_{0}$ from here on.
The second factor carries the attenuation generated by $-\frac{i}{2}\gamma_{+}\hat{n}_{+}$.
The third factor gives the reduced dimer evolution generated by Eq.~\eqref{eq:Hred}.
In the rotating frame, the conditioned no-jump state evolves as
\begin{align}
    \lvert \tilde{\psi}(t) \rangle =&~ e^{- \frac{1}{2} \gamma_{+} \hat{n}_{+} t } e^{-\frac{i}{\hbar} \hat{H}_{n_{+}} t } \lvert \tilde{\psi}(0) \rangle.
\end{align}

The common-mode attenuation factor does not, in general, coincide with the full no-jump probability. Decomposing the initial state into total-excitation sectors, $\lvert\tilde{\psi}(0)\rangle=\sum_{N}\lvert\psi_{N}(0)\rangle$, with $\hat{n}_{+}\lvert\psi_{N}(0)\rangle=N\lvert\psi_{N}(0)\rangle$, we define the reduced propagator within the $N$-excitation sector as
\begin{align}
    \hat{U}_{N}(t) =&~
    e^{-\frac{i}{\hbar}\hat{H}_{n_{+}}^{(N)}t} \,,
\end{align}
where $\hat{H}_{n_{+}}^{(N)}$ denotes the restriction of $\hat{H}_{n_{+}}$ to that sector, after the common $-i\gamma_{+}\hat{n}_{+}/2$ contribution has been separated. The no-jump probability then becomes
\begin{align}
    P_{0}(t) =&~ \sum_{N} e^{-\gamma_{+}Nt} \left\langle\psi_{N}(0)\right| \hat{U}_{N}^{\dagger}(t)\hat{U}_{N}(t) \left|\psi_{N}(0)\right\rangle \, .
\end{align}
The rate $\gamma_{+}$ sets the common excitation-number-dependent
attenuation scale, while the full no-jump probability also records the
differential-loss dynamics through $\hat{U}_{N}(t)$. When
$\gamma_{-}=0$, the reduced Hamiltonian is Hermitian and
$\hat{U}_{N}(t)$ is unitary. A normalized fixed-$N$ initial state then
yields $P_{0}(t)=e^{-\gamma_{+}Nt}$, whereas an initial state spanning
several excitation-number sectors retains the corresponding weighted
sum over $N$.

Within the $N$-excitation sector, the loss operator
$\gamma_{1}\hat{n}_{1}+\gamma_{2}\hat{n}_{2}$ has eigenvalues between
$\gamma_{\min}N$ and $\gamma_{\max}N$, where
$\gamma_{\min}=\gamma_{+}-\lvert\gamma_{-}\rvert$ and
$\gamma_{\max}=\gamma_{+}+\lvert\gamma_{-}\rvert$.
Applying the preceding norm-loss equation separately to each excitation
sector therefore gives
\begin{align}
    \sum_{N}p_{N}e^{-\gamma_{\max}Nt}
    \le P_{0}(t) \le
    \sum_{N}p_{N}e^{-\gamma_{\min}Nt},
\end{align}
where
$p_{N}=\langle\psi_{N}(0)\vert\psi_{N}(0)\rangle$ and
$\sum_{N}p_{N}=1$ for a normalized initial state.
When $\gamma_{-}=0$, the two bounds coincide and yield the exact
weighted sum $\sum_{N}p_{N}e^{-\gamma_{+}Nt}$.
For $\gamma_{-}\neq0$, the common-mode attenuation alone does not,
in general, determine $P_{0}(t)$.

For detuned modes,
$\omega_{1,2} = \omega_{0} \pm \Delta$ with $\Delta
= \left( \omega_{1} - \omega_{2} \right)/2$,
the free Hamiltonian adds the differential term $\Delta \hat{n}_{-}$ to the reduced dimer.
The coefficient of $\hat{n}_{-}$ becomes $\Delta - \frac{i}{2}\gamma_{-}$, and the squared reduced eigenvalue scale becomes
\begin{align}
    \label{eq:KappaDetuned}
    \kappa^{2} = \frac{1}{4}\gamma_{-}^{2} - g^{2} - \Delta^{2} + i \gamma_{-} \Delta.
\end{align}
The reduced spectrum is generically complex, so the conditioned flow combines rotation and stretching instead of falling into a purely elliptic, parabolic, or hyperbolic class.
An exceptional point requires $\gamma_{-}\Delta = 0$ together with $\gamma_{-}^{2} = 4\left( g^{2} + \Delta^{2} \right)$, which forces $\Delta = 0$ for nonzero hopping in this parametrization.
For this reason, we restrict the main analysis to resonant modes, where the non-Hermitian dimer has a real regime parameter and a sharp transition between elliptic, parabolic, and hyperbolic flow.

For vanishing decay rates $\gamma_{j} = 0$, we recover the lossless bosonic dimer.
Choosing $\phi = \pi/2$ gives the standard lossless beam-splitter Hamiltonian, which realizes a $50{:}50$ beam splitter at the interaction time $gt = \pi/4$.

\section{Grid state evolution}
\label{sec:Sec3}

We describe the conditioned evolution in phase space through dimensionless quadratures,
\begin{align}
    \begin{aligned}
        \hat{q}_{j} =&~ \frac{1}{\sqrt{2}} \left( \hat{a}_{j} + \hat{a}_{j}^{\dagger} \right), \\
        \hat{p}_{j} =&~ \frac{-i}{\sqrt{2}} \left( \hat{a}_{j} - \hat{a}_{j}^{\dagger} \right),
    \end{aligned}
\end{align}
ordered in the phase-space vector
\begin{align}
    \hat{\bm{\chi}} =&~ \left( \hat{q}_{1}, \hat{q}_{2}, \hat{p}_{1}, \hat{p}_{2} \right)^{\mathrm{T}}.
\end{align}
The canonical commutation relations take the compact form
\begin{align}
    \left[ \hat{\chi}_{j}, \hat{\chi}_{k} \right] =&~ i \Omega_{jk},
\end{align}
with symplectic form
\begin{align}
    \bm{\Omega} =&~
    \begin{pmatrix}
        \bm{0}_{2} & \bm{I}_{2} \\
        -\bm{I}_{2} & \bm{0}_{2}
    \end{pmatrix}.
\end{align}
A phase-space vector $\bm{v} \in \mathbb{R}^{4}$ defines the displacement operator
\begin{align}
    \hat{D}\left( \bm{v} \right) =&~ e^{-i \bm{v}^{\mathrm{T}}\bm{\Omega}\hat{\bm{\chi}}}.
\end{align}
The initial grid is generated by real displacement vectors, for which
$\hat{D}\left(\bm{v}\right)$ is an ordinary unitary Weyl displacement.
Under non-Hermitian evolution, however, the displacement parameters generally
become complex. Writing
$\bm{v}=\bm{a}+i\bm{b}$, with
$\bm{a},\bm{b}\in\mathbb{R}^{4}$, the exponent in
\begin{align}
    \hat{D}\left(\bm{v}\right)
    =e^{-i\bm{v}^{\mathrm{T}}\bm{\Omega}\hat{\bm{\chi}}}
\end{align}
separates as
\begin{align}
    -i\bm{v}^{\mathrm{T}}\bm{\Omega}\hat{\bm{\chi}}
    = -i\bm{a}^{\mathrm{T}}\bm{\Omega}\hat{\bm{\chi}}
    +\bm{b}^{\mathrm{T}}\bm{\Omega}\hat{\bm{\chi}}.
\end{align}
Although these two terms generally do not commute, their commutator is a scalar. Defining
\begin{align}
    \hat{X} =&~ \bm{b}^{\mathrm{T}}\bm{\Omega}\hat{\bm{\chi}} \, \,,
    \quad \hat{Y} = -i\bm{a}^{\mathrm{T}}\bm{\Omega}\hat{\bm{\chi}} \,,
\end{align}
the canonical commutation relations imply
\begin{align}
    [\hat{X},\hat{Y}] =&~
    -i\left[
    \bm{b}^{\mathrm{T}}\bm{\Omega}\hat{\bm{\chi}},
    \bm{a}^{\mathrm{T}}\bm{\Omega}\hat{\bm{\chi}}
    \right] \nonumber\\
    =&~ \bm{b}^{\mathrm{T}}\bm{\Omega}\bm{a}
    = -\bm{a}^{\mathrm{T}}\bm{\Omega}\bm{b}.
\end{align}
Because this commutator is a scalar, all higher nested commutators vanish.
The Baker--Campbell--Hausdorff formula therefore gives
\begin{align}
    \hat{D}\left(\bm{a}+i\bm{b}\right)
    =&~
    e^{\hat{X}+\hat{Y}}
    \nonumber\\
    =&~
    e^{-\frac{1}{2}[\hat{X},\hat{Y}]}
    e^{\hat{X}}e^{\hat{Y}}
    \nonumber\\
    =&~
    e^{\frac{1}{2}\bm{a}^{\mathrm{T}}\bm{\Omega}\bm{b}}\,
    e^{\bm{b}^{\mathrm{T}}\bm{\Omega}\hat{\bm{\chi}}}\,
    \hat{D}\left(\bm{a}\right).
\end{align}

The ``complex lattice'' considered below therefore denotes a lattice of
complex parameters in $\mathbb{C}^4$ that label analytically continued
displacement operators. It does not represent a lattice of directly observable
points in the real phase space $\mathbb{R}^4$. When the transformed
displacement vectors remain real, the nonunitary reweighting operator is absent,
the operators reduce to ordinary unitary Weyl displacements, and the lattice
retains its conventional interpretation as a real phase-space GKP lattice.

Within this phase-space formalism, a two-mode displaced comb in configuration space,
\begin{align}
    \label{eq:IdealGrid}
    \Psi_{\mathrm{grid}}(q_{1}, q_{2}) =&~ \sum_{m,n\in\mathbb{Z}} \hat{D}\left( m\bm{u}_{1} + n\bm{u}_{2} \right) \delta(q_{1}) \delta(q_{2}),
\end{align}
displaces a singular origin seed over the lattice generated by the primitive vectors $\bm{u}_{1}, \bm{u}_{2} \in \mathbb{R}^{4}$ with linearly independent configuration projections.
This ideal state is unphysical because the Dirac-delta seed is not normalizable.

Under the conditioned bosonic-dimer evolution, the grid state
\begin{align}
    \Psi_{\mathrm{grid}}(q_{1}, q_{2},t) =&~ \sum_{m,n\in\mathbb{Z}} \hat{\mathcal{D}}_{m,n}\left( t \right)\Phi_{0}\left( q_{1}, q_{2},t \right),
\end{align}
separates into the evolution of the displacement lattice,
\begin{align}
    \hat{\mathcal{D}}_{m,n}\left( t \right) =&~ e^{-\frac{1}{2}\gamma_{+}\hat{n}_{+}t} e^{-\frac{i}{\hbar} \hat{H}_{n_{+}} t } \hat{D}\left( m\bm{u}_{1} + n\bm{u}_{2} \right) \nonumber \\
    &~ \times e^{\frac{i}{\hbar} \hat{H}_{n_{+}} t } e^{\frac{1}{2}\gamma_{+}\hat{n}_{+}t},
\end{align}
and the evolution of the origin seed,
\begin{align}
    \begin{aligned}
        \Phi_{0}\left( q_{1}, q_{2},t \right) =&~ \lim_{\sigma \to 0} \frac{1}{2 \sqrt{\pi} \sigma} \Phi_{\sigma}\left( q_{1}, q_{2},t \right), \\
        \Phi_{\sigma}\left( q_{1}, q_{2},t \right) =&~ \left\langle q_{1}, q_{2} \middle| e^{- \frac{1}{2} \gamma_{+} \hat{n}_{+} t } e^{-\frac{i}{\hbar} \hat{H}_{n_{+}} t } \middle| \Phi_{\sigma}(0) \right\rangle, \\
        \Phi_{\sigma}\left( q_{1}, q_{2},0 \right) =&~ \frac{1}{\sqrt{\pi}\,\sigma} e^{- \frac{1}{2 \sigma^{2}} \left( q_{1}^{2} + q_{2}^{2} \right)}.
    \end{aligned}
\end{align}
Here $\lvert\Phi_{\sigma}(0)\rangle$ denotes the normalized Gaussian state
whose configuration-space wavefunction is given in the last line above.
The Dirac-delta seed is obtained by taking the zero-width limit of this
Gaussian after applying the indicated rescaling factor. This rescaling is
necessary because, although $\Phi_{\sigma}(q_{1},q_{2},0)$ has unit
$L^{2}$ norm, its integral is proportional to $\sigma$ and therefore
vanishes as $\sigma\to0$. The rescaled Gaussian instead converges
distributionally to $\delta(q_{1})\delta(q_{2})$. For finite $\sigma$, this
construction regularizes each lattice peak and allows its evolution to be
described using the Siegel parametrization of Gaussian states.
A physically finite-energy grid state additionally requires
a decaying envelope over the lattice.

We use the conditioned propagator,
\begin{align}
    \label{eq:ConditionedPropagator}
    \hat{U}(t) = e^{-\frac{1}{2}\gamma_{+}\hat{n}_{+}t}
    e^{-\frac{i}{\hbar}\hat{H}_{n_{+}}t} \,,
\end{align}
to construct the quadrature similarity transformation that induces the
evolution of the lattice vectors,
\begin{align}
    \hat{U}^{-1}(t)\hat{\bm{\chi}}\hat{U}(t)
    = \bm{S}(t)\hat{\bm{\chi}} \,,
\end{align}
where
\begin{align} \label{smatrix}
    \bm{S}(t)
    = e^{-\frac{i}{2}\gamma_{+}t\bm{\Omega}}
    e^{t\bm{K}_{n_{+}}} \, .
\end{align}
Here, the reduced-dimer factor is generated by
\begin{align}
    \bm{K}_{n_{+}}
    = \begin{pmatrix}
    0 & -g\sin\phi & -\frac{i}{2}\gamma_{-} & g\cos\phi \\
    g\sin\phi & 0 & g\cos\phi & \frac{i}{2}\gamma_{-} \\
    \frac{i}{2}\gamma_{-} & -g\cos\phi & 0 & -g\sin\phi \\
    -g\cos\phi & -\frac{i}{2}\gamma_{-} & g\sin\phi & 0
\end{pmatrix} \, .
\end{align}
This generator satisfies
$\bm{K}_{n_{+}}^{\mathrm{T}}\bm{\Omega}
+\bm{\Omega}\bm{K}_{n_{+}} = 0$, and therefore belongs to
the complex symplectic Lie algebra $\mathrm{sp}(4,\mathbb{C})$.
Its propagator $e^{t\bm{K}_{n_{+}}}$ satisfies
$e^{t\bm{K}_{n_{+}}^{\mathrm{T}}}\bm{\Omega}e^{t\bm{K}_{n_{+}}} = \bm{\Omega}$,
and thus belongs to the complex symplectic group
$\mathrm{Sp}(4,\mathbb{C})$.

The matrix generator
$\bm{A}_{+}=-\frac{i}{2}\gamma_{+}\bm{\Omega}$ of the first factor in
$\bm{S}(t)$ also satisfies the symplectic Lie-algebra condition
$\bm{A}_{+}^{\mathrm{T}}\bm{\Omega}
+\bm{\Omega}\bm{A}_{+}=0$.
Therefore,
$e^{t\bm{A}_{+}}
=e^{-\frac{i}{2}\gamma_{+}t\bm{\Omega}}$
belongs to $\mathrm{Sp}(4,\mathbb{C})$. Since the product of complex
symplectic matrices is complex symplectic, the full similarity matrix
$\bm{S}(t)$ satisfies
\begin{align}
    \bm{S}^{\mathrm{T}}(t)\bm{\Omega}\bm{S}(t)
    = \bm{\Omega}.
\end{align}
The similarity matrix $\bm{S}(t)$ transports displacement
parameters under the non-unitary propagator $\hat{U}(t)$.
It should be distinguished from
the adjoint transformation involving $\hat{U}^{\dagger}(t)$ that enters
normalized conditional expectation values. From
\begin{align}
    \hat{U}(t)\hat{\bm{\chi}}\hat{U}^{-1}(t)
    = \bm{S}^{-1}(t)\hat{\bm{\chi}}
\end{align}
and the symplectic identity
$\bm{\Omega}\bm{S}^{-1}(t)=\bm{S}^{\mathrm{T}}(t)\bm{\Omega}$, we obtain
\begin{align}
    \hat{U}(t)\hat{D}(\bm{v})\hat{U}^{-1}(t)
    =& \exp\left[-i\bm{v}^{\mathrm{T}}\bm{\Omega}
    \bm{S}^{-1}(t)\hat{\bm{\chi}} \right]
    \nonumber\\
    =& \hat{D}\left[\bm{S}(t)\bm{v}\right] \, .
\end{align}
Therefore, for the lattice vector
$\bm{v}_{m,n}=m\bm{u}_{1}+n\bm{u}_{2}$, the conditioned bosonic-dimer
evolution gives
\begin{align}
    \hat{\mathcal{D}}_{m,n}(t)
    = \hat{D}\left[
    e^{-\frac{i}{2}\gamma_{+}t\bm{\Omega}}
    e^{t\bm{K}_{n_{+}}}
    \left(m\bm{u}_{1}+n\bm{u}_{2}\right)
    \right] \, .
\end{align}

The complex-symplectic character of this deformation also
identifies which properties of the GKP lattice remain invariant under the
conditioned dynamics.
The symplectic product of any two lattice vectors transformed by the matrix
$\bm{S}(t)$ defined in \eqref{smatrix} remains invariant,
\begin{align}
    \left[\bm{S}(t) \bm{v}\right]^{\mathrm{T}}
    \bm{\Omega}
    \left[\bm{S}(t) \bm{w}\right]
    = \bm{v}^{\mathrm{T}}\bm{\Omega}\bm{w} \,,
\end{align}
which implies that their displacement operators retain the same
algebraic commutation phase,
\begin{align} \label{algcp}
    \hat{D}(\bm{S}(t)\bm{v})\hat{D}(\bm{S}(t)\bm{w})
    = e^{-i\bm{v}^{\mathrm{T}}\bm{\Omega}\bm{w}}\hat{D}(\bm{S}(t)\bm{w})\hat{D}(\bm{S}(t)\bm{v}).
\end{align}
For GKP encodings, these symplectic products determine the commutation relations among stabilizer and logical displacement operators \cite{Gottesman2001p016316,Mensen2021p022408,Royer2022p010335}.

The preservation of the displacement-operator commutation relation in
\eqref{algcp} does not imply that the conditioned state remains a
conventional GKP error-correcting code. In the lossless limit,
$\gamma_{1}=\gamma_{2}=0$, the transformation is real symplectic and transforms
the initial grid into another real phase-space grid generated by unitary Weyl
displacements. For nonzero conditioned loss, the transformed lattice vectors
generally become complex, and the associated operators become non-unitary
rather than ordinary physical Weyl displacements. The complex lattice retains
the symplectic displacement algebra but does not, in general, define a standard
real phase-space GKP stabilizer code to which the usual displacement-syndrome
and correctable-region interpretation applies directly. For a finite-energy
grid state, the simultaneous evolution of the regularized seed also
modifies the local Gaussian peak shape and, therefore, the state's response
to displacement errors.
A quantitative assessment of logical error-correction performance requires a
specified physical noise model, syndrome measurement, and recovery operation
and lies beyond the scope of the present work.

We now evaluate the regularized origin-seed evolution.
The quadrature similarity matrix has the complex symplectic block form
\begin{align}
    \bm{S}(t) = e^{-\frac{i}{2} \gamma_{+} t \bm{\Omega}} e^{t\bm{K}_{n_{+}}} =&~
    \begin{pmatrix}
        \bm{A}(t) & \bm{B}(t) \\
        -\bm{B}(t) & \bm{A}(t)
    \end{pmatrix}.
\end{align}
Up to the scalar non-unitary normalization factor that controls the no-jump weight, the evolved regularized seed keeps the quadratic form
\begin{align}
    \Phi_{\sigma}\left( \bm{q}, t \right) \propto&~
    e^{ \frac{i}{2}\bm{q}^{\mathrm{T}}\bm{\Gamma}_{\sigma}(t)\bm{q}},
\end{align}
with $\bm{q} = \left( q_{1}, q_{2} \right)^{\mathrm{T}}$ and complex width matrix
\begin{align}
    \bm{\Gamma}_{\sigma}(t) =&~ \frac{i}{\sigma^{2}}
    \left[ \bm{A}(t) + i \sigma^{2} \bm{B}(t) \right]
    \left[ \bm{A}(t) + \frac{i}{\sigma^{2}} \bm{B}(t) \right]^{-1},
\end{align}
which recovers $\bm{\Gamma}_{\sigma}(0) = i \sigma^{-2} \bm{I}_{2}$.
Writing
\begin{align}
    \bm{\Gamma}_{\sigma}(t) =&~ \bm{X}_{\sigma}(t) + i \bm{Y}_{\sigma}(t),
\end{align}
we obtain the normalized Gaussian shape
\begin{align}
    \Phi_{\sigma}^{(\mathrm{sh})}\left( \bm{q}, t \right) =&~
    \mathcal{N}_{\sigma}(t)
    e^{ \frac{i}{2}\bm{q}^{\mathrm{T}}\bm{X}_{\sigma}(t)\bm{q}}
    e^{ - \frac{1}{2}\bm{q}^{\mathrm{T}}\bm{Y}_{\sigma}(t)\bm{q}},
\end{align}
where $\mathcal{N}_{\sigma}(t)$ provides unit $L^{2}$ norm.
The imaginary part, $\bm{Y}_{\sigma}(t)$, determines the principal axes and widths of the seed density, whose configuration covariance is $[2\bm{Y}_{\sigma}(t)]^{-1}$, while the real part, $\bm{X}_{\sigma}(t)$, sets its quadratic phase.
Common-mode attenuation also contributes to the seed evolution, alongside coherent hopping and differential loss.
For the beam-splitter phase, $\phi=\pi/2$, the real part vanishes,
$\bm{X}_{\sigma}(t)=\bm{0}_{2}$.
The evolved regularized seed defines a square-integrable Gaussian shape only when $\bm{Y}_{\sigma}(t)$ is positive definite,
\begin{align}
    \begin{aligned}
        \operatorname{tr}\bm{Y}_{\sigma}(t) >&~ 0, \\
        \det\bm{Y}_{\sigma}(t) >&~ 0.
    \end{aligned}
\end{align}

\section{Grid dynamics}
\label{sec:Sec4}

At the phase-space level, common-mode attenuation induces the complex symplectic transformation
\begin{align}
    e^{-\frac{i}{2}\gamma_{+}t\bm{\Omega}} =&~ \cosh\left( \frac{\gamma_{+}t}{2} \right)\bm{I}_{4} - i\sinh\left( \frac{\gamma_{+}t}{2} \right)\bm{\Omega},
\end{align}
since $\bm{\Omega}^{2} = -\bm{I}_{4}$.
This factor mixes configuration and momentum directions through a complex symplectic rotation.

After separating the common-mode factor, the reduced
conditioned dynamics makes the distinct roles of coherent hopping and
differential loss transparent. The hopping $g$ generates beam-splitter-like
mixing between the two modes and, by itself, rotates and transfers lattice
directions between them. Differential loss instead produces unequal no-jump
backaction on the two modes. At fixed total excitation number, if
$\gamma_{1}>\gamma_{2}$, components with greater occupation of mode $1$ are
more strongly suppressed than components with greater occupation of mode $2$,
with the roles reversed when $\gamma_{2}>\gamma_{1}$. Because coherent
hopping continually redistributes excitation between the modes, this unequal
conditional reweighting becomes a directional deformation of the lattice
vectors. The competition between $g$ and $\gamma_{-}$ controls the reduced
dynamical regimes: bounded oscillatory deformation in the elliptic regime,
linear deformation at the exceptional point, and exponentially separating
lattice directions in the hyperbolic regime. The complex components produced
by the conditioned-loss terms encode nonunitary conditional reweighting and
do not represent additional physical phase-space coordinates.

The reduced dimer factor controls the regime-dependent deformation of the primitive lattice vectors.
Its generator satisfies
\begin{align}
    \bm{K}_{n_{+}}^{2} =&~ \kappa^{2} \bm{I}_{4},
\end{align}
with
\begin{align}
    \kappa =&~ \frac{1}{2}\sqrt{\gamma_{-}^{2} - 4g^{2}},
\end{align}
and has the doubly degenerate spectrum
\begin{align}
    \mathrm{spec}\left( \bm{K}_{n_{+}} \right) =&~ \left\{ \kappa, \kappa, -\kappa, -\kappa \right\}.
\end{align}
This spectrum separates the reduced flow of the resonant dimer into three regimes.
For $\gamma_{-}^{2} < 4g^{2}$, the scale $\kappa$ is imaginary and the reduced flow is elliptic.
For $\gamma_{-}^{2} > 4g^{2}$, the scale $\kappa$ is real and the reduced flow is hyperbolic.
At $\gamma_{-}^{2} = 4g^{2}$, the scale vanishes, the generator becomes nilpotent and non-diagonalizable, $\bm{K}_{n_{+}}^{2} = \bm{0}_{4}$ with $\bm{K}_{n_{+}} \neq \bm{0}_{4}$, and the critical point is an exceptional point.

The same quadratic identity gives the reduced propagator,
\begin{align}
    e^{t \bm{K}_{n_{+}}} =&~
    \begin{cases}
        e^{\kappa t} \bm{\Pi}_{+} + e^{-\kappa t} \bm{\Pi}_{-}, & \kappa \neq 0, \\
        \bm{I}_{4} + t \bm{K}_{n_{+}}, & \kappa = 0,
    \end{cases}
\end{align}
where the complementary projectors
\begin{align}
    \bm{\Pi}_{\pm} =&~ \frac{1}{2} \left( \bm{I}_{4} \pm \frac{1}{\kappa} \bm{K}_{n_{+}} \right),
\end{align}
satisfy $\bm{\Pi}_{\pm}^{2} = \bm{\Pi}_{\pm}$, $\bm{\Pi}_{+}\bm{\Pi}_{-} = \bm{0}_{4}$, $\bm{\Pi}_{+} + \bm{\Pi}_{-} = \bm{I}_{4}$, and $\bm{\Pi}_{+} - \bm{\Pi}_{-} = \kappa^{-1} \bm{K}_{n_{+}}$ for $\kappa \neq 0$.

In the elliptic regime, $\kappa = i\lvert \kappa \rvert$, the reduced flow has period $\lvert \kappa \rvert T = 2\pi$.
The reduced evolution returns the input lattice up to a sign at
\begin{align}
    \begin{aligned}
        e^{t_{m}^{(\mathrm{rev})}\bm{K}_{n_{+}}} =&~ \left( -1 \right)^{m}\bm{I}_{4}, \\
        t_{m}^{(\mathrm{rev})} =&~ \frac{m\pi}{\lvert \kappa \rvert},
    \end{aligned}
\end{align}
with $m \in \mathbb{Z}_{\geq 0}$, and has a vanishing identity coefficient at
\begin{align}
    \begin{aligned}
        e^{t_{m}^{(0)}\bm{K}_{n_{+}}} =&~ \left( -1 \right)^{m}\frac{1}{\lvert \kappa \rvert}\bm{K}_{n_{+}}, \\
        t_{m}^{(0)} =&~ \frac{\pi}{\lvert \kappa \rvert}\left( m + \frac{1}{2} \right).
    \end{aligned}
\end{align}
Between these limits, the beam-splitter-like times
\begin{align}
    \begin{aligned}
        e^{t_{m}^{(\mathrm{bs},+)}\bm{K}_{n_{+}}} =&~ \frac{\left( -1 \right)^{m}}{\sqrt{2}}
        \left[ \bm{I}_{4} + \frac{1}{\lvert \kappa \rvert}\bm{K}_{n_{+}} \right], \\
        t_{m}^{(\mathrm{bs},+)} =&~ \frac{\pi}{\lvert \kappa \rvert}\left( m + \frac{1}{4} \right),
    \end{aligned}
\end{align}
and
\begin{align}
    \begin{aligned}
        e^{t_{m}^{(\mathrm{bs},-)}\bm{K}_{n_{+}}} =&~ -\frac{\left( -1 \right)^{m}}{\sqrt{2}}
        \left[ \bm{I}_{4} - \frac{1}{\lvert \kappa \rvert}\bm{K}_{n_{+}} \right], \\
        t_{m}^{(\mathrm{bs},-)} =&~ \frac{\pi}{\lvert \kappa \rvert}\left( m + \frac{3}{4} \right),
    \end{aligned}
\end{align}
give equal magnitudes for the identity and
$\bm{K}_{n_{+}}/\lvert\kappa\rvert$ coefficients.
For identical losses, $\gamma_{-}=0$, normalized evolution within each fixed total-excitation sector coincides with the lossless beam splitter and is balanced at $gt=\pi/4$.
Common-mode attenuation still reweights a grid state occupying several sectors.
For one excitation initially in mode 1, the transferred no-jump weight in mode 2 is $e^{-\gamma_{+}t}\, g^{2}\sin^{2}
\left( \lvert\kappa\rvert t \right)/\lvert\kappa\rvert^{2}$, with its first maximum at
$\cot\left( \lvert\kappa\rvert t \right) =
\gamma_{+}/(2\lvert\kappa\rvert)$.

In the hyperbolic regime, $\kappa > 0$, the reduced flow has no finite period.
The complementary projectors isolate the two invariant sectors,
\begin{align}
    e^{t\bm{K}_{n_{+}}}\bm{\Pi}_{\pm}
    = e^{\pm\kappa t}\bm{\Pi}_{\pm}.
\end{align}
Thus, for $t\ge0$, the $\bm{\Pi}_{-}$ sector decays exponentially,
whereas the $\bm{\Pi}_{+}$ sector grows exponentially in the reduced
frame.
This exponential behavior belongs to the reduced frame because we separated the common-mode attenuation through $e^{- \frac{1}{2}\gamma_{+}\hat{n}_{+}t}$; it does not imply amplification of the full conditioned state.

At the critical point, $\kappa = 0$, the reduced flow becomes parabolic.
The propagator truncates to
\begin{align}
    e^{t\bm{K}_{n_{+}}} =&~ \bm{I}_{4} + t\bm{K}_{n_{+}},
\end{align}
with asymptotic renormalized limit
\begin{align}
    \lim_{t\to\infty} \frac{1}{t} e^{t\bm{K}_{n_{+}}} =&~ \bm{K}_{n_{+}}.
\end{align}

The significance of the three regimes is that they characterize the grid deformation produced by no-jump conditioning in a lossy bosonic dimer.
The unconditioned dissipative Gaussian channel describes the ensemble of all outcomes, while post-selecting trajectories without loss events, for example, gives this conditioned non-Hermitian evolution.
The balance between differential loss and coherent hopping determines the reduced displacement flow; it is bounded and oscillatory below the exceptional point, linear at the exceptional point, and exponential above it.
Our contribution is the explicit connection between these operating regimes and the deformation of ideal grid states.
In the lossy bosonic dimer, conditioning dissipative Gaussian dynamics on the no-jump sector produces the grid deformations characterized by these three reduced displacement flows.
Such critical behavior arising from eigenvalue and eigenvector coalescence is characteristic of exceptional-point physics \cite{Ashida2020p015005,Miri2019peaar7709}.

\section{Seed dynamics}
\label{sec:Sec5}

The complex width matrix $\bm{\Gamma}_{\sigma}(t)$ controls the
regularized seed evolution.
Its deformation is governed by coherent hopping and by the common-mode and
differential components of the conditioned loss.
Whereas the lattice vectors label the repeated displaced components of
the grid and determine their phase-space positions when the vectors are
real, the seed determines the shape of each individual peak. In
$\bm{\Gamma}_{\sigma}(t)=\bm{X}_{\sigma}(t)+i\bm{Y}_{\sigma}(t)$,
the matrix $\bm{Y}_{\sigma}(t)$ determines the Gaussian envelope,
including its principal widths and directions, while
$\bm{X}_{\sigma}(t)$ determines the quadratic phase across the peak.
Coherent hopping mixes the quadrature directions of the two modes, common-mode
attenuation modifies the overall Gaussian width, and differential loss
produces mode-dependent reweighting through the no-jump backaction. Their
combined action can therefore turn the initially isotropic seed into an
anisotropic, correlated Gaussian with a nontrivial quadratic phase.

The complex width matrix $\bm{\Gamma}_{\sigma}(t)$ is well defined whenever the denominator
\begin{align}
    \Delta_{\sigma}\left( t \right) =&~ \left[ \sigma^{2}C_{+}(t)+S_{+}(t) \right]^{2}
    +\left(1-\sigma^{4}\right)\left[ g^{2}\cos^{2}(\phi)-\frac{\gamma_{-}^{2}}{4} \right]S^{2}(t)
    \label{eq:SeedDenominator}
\end{align}
does not vanish.
We use the shorthand $C_{+}\left( t \right) = \cosh\left( \gamma_{+}t/2 \right)$, $S_{+}\left( t \right) = \sinh\left( \gamma_{+}t/2 \right)$, $C\left( t \right) = \cosh\left( \kappa t \right)$, and $S\left( t \right) = \sinh\left( \kappa t \right)/\kappa$, with the elliptic regime obtained by analytic continuation to $\kappa \in i\mathbb{R}$ and the parabolic regime obtained from the limit $\kappa \to 0$.

First, we show that the denominator is strictly positive for every physical Gaussian waist $\sigma > 0$, evolution time $t \ge 0$, and loss rate $\gamma_{+} \ge \lvert \gamma_{-} \rvert$.
In the elliptic and parabolic regimes, $\lvert S(t)\rvert\le t$.
In the hyperbolic regime, $0<\kappa\le\lvert\gamma_{-}\rvert/2$ and $\sinh(\kappa t)/\kappa$ is non-decreasing in $\kappa$.
The differential loss rate bounds the magnitude of $S(t)$ across all three regimes,
\begin{align} \label{eq:SeedDimerBound}
    \lvert S(t) \rvert
    \le t\operatorname{sinhc}\left(
    \frac{1}{2}\lvert\gamma_{-}\rvert t
    \right),
\end{align}
where $\operatorname{sinhc}(z)=\sinh(z)/z$ for $z\ne0$ and $\operatorname{sinhc}(0)=1$.
This bound and the identity $C^{2}(t)-\kappa^{2}S^{2}(t)=1$ restrict the coefficient in $\Delta_{\sigma}(t)$ to
\begin{align} \label{eq:SeedCoefficientBound}
    -S_{+}^{2}(t) \le
    \left[ g^{2}\cos^{2}(\phi)-\frac{\gamma_{-}^{2}}{4} \right]S^{2}(t) \le 1 \, .
\end{align}
The sign of $1-\sigma^{4}$ in $\Delta_{\sigma}(t)$ selects the lower coefficient bound for $0<\sigma\le1$ and the upper one for $\sigma\ge1$,
\begin{align}
    \begin{aligned}
        \Delta_{\sigma}(t) \ge &~ \begin{cases}
        \sigma^{4}\left[ C_{+}^{2}(t)+S_{+}^{2}(t) \right]+2\sigma^{2}C_{+}(t)S_{+}(t) \ge \sigma^{4}, & 0<\sigma\le1, \\[1mm]
        \left[ \sigma^{2}C_{+}(t)+S_{+}(t) \right]^{2}+1-\sigma^{4} \ge 1, & \sigma\ge1,
    \end{cases} \\[1mm]
    \ge &~ \min\{\sigma^{4},1\}>0.
\end{aligned}
\label{eq:SeedDenominatorBound}
\end{align}
The denominator is strictly positive throughout the physical domain, which establishes that $\bm{\Gamma}_{\sigma}(t)$ is well defined.

The conditions $C_{+}(t)\ge1$ and $S_{+}(t)\ge0$ make both factors in
\begin{align} \label{eq:SeedUpsilon}
    \Upsilon_{\sigma}(t)
    =\left[\sigma^{2}C_{+}(t)+S_{+}(t)\right]
    \left[C_{+}(t)+\sigma^{2}S_{+}(t)\right]
\end{align}
strictly positive for $\sigma>0$.
Division by the positive denominator preserves this sign in the trace,
\begin{align}
    \operatorname{tr}\bm{Y}_{\sigma}(t)
    = \frac{2\Upsilon_{\sigma}(t)}{\Delta_{\sigma}(t)} > 0 \, .
    \label{eq:SeedTrace}
\end{align}

In the determinant,
\begin{align}
    \det\bm{Y}_{\sigma}(t) =&~ \frac{\Upsilon_{\sigma}^{2}(t)-\lambda_{\sigma}^{2}\gamma_{-}^{2}S^{2}(t)\left[C^{2}(t)+g^{2}S^{2}(t)\sin^{2}(\phi)\right]}{\Delta_{\sigma}^{2}(t)},
    \label{eq:SeedDeterminant}
\end{align}
with $\lambda_{\sigma}=(\sigma^{4}-1)/2$,
differential loss subtracts a non-negative term
from $\Upsilon_{\sigma}^{2}(t)$ in the numerator.
We use the identity $C^{2}(t)-\kappa^{2}S^{2}(t)=1$ to isolate the non-positive hopping contribution,
\begin{align}
    C^{2}(t)+g^{2}S^{2}(t)\sin^{2}(\phi) &=
    1+\left[\frac{\gamma_{-}^{2}}{4}-g^{2}\cos^{2}(\phi)\right]S^{2}(t) \nonumber \\[1mm]
    & \le 1+\frac{\gamma_{-}^{2}}{4}S^{2}(t)\le\cosh^{2}\left(\frac{1}{2}\lvert\gamma_{-}\rvert t\right),
    \label{eq:SeedSquareRootBound}
\end{align}
where the final inequality uses the bound on $S(t)$.
After taking the square root and multiplying by $\lvert\gamma_{-}S(t)\rvert$, the bounds combine through the double-angle identity,
\begin{align} \label{eq:SeedLossBound}
    \lvert \gamma_{-}S(t) \rvert\sqrt{C^{2}(t)+g^{2}S^{2}(t)\sin^{2}(\phi)} \le &~ 2\sinh\left(\frac{1}{2}\lvert\gamma_{-}\rvert t\right)\cosh\left(\frac{1}{2}\lvert\gamma_{-}\rvert t\right) \nonumber \\
    = &~ \sinh\left(\lvert\gamma_{-}\rvert t\right) \le \sinh(\gamma_{+}t),
\end{align}
where physical losses enforce the last inequality.
In the expansion of $\Upsilon_{\sigma}(t)$, the coefficient of $\sinh(\gamma_{+}t)$ exceeds $\lvert\lambda_{\sigma}\rvert$ by $(1+\sigma^{4})/2-\lvert\lambda_{\sigma}\rvert=\min\{\sigma^{4},1\}>0$.
Subtracting the bounded differential-loss contribution
leaves a strictly positive difference,
\begin{align}
    & \Upsilon_{\sigma}(t)-\lvert\lambda_{\sigma}\gamma_{-}S(t)\rvert
    \sqrt{C^{2}(t)+g^{2}S^{2}(t)\sin^{2}(\phi)} \nonumber \\[1mm]
    & \ge \sigma^{2}\cosh(\gamma_{+}t)+\min\{\sigma^{4},1\}\sinh(\gamma_{+}t) \nonumber \\[1mm]
    & \ge \sigma^{2}\cosh(\gamma_{+}t) > 0 \, . \label{eq:SeedPositivityBound}
\end{align}
The determinant numerator is the product of this positive difference and the corresponding sum, which is at least as large.
Both factors are bounded below by $\sigma^{2}\cosh(\gamma_{+}t)$, making the determinant strictly positive,
\begin{align} \label{eq:SeedDeterminantBound}
    \det\bm{Y}_{\sigma}(t) \ge \frac{\sigma^{4}\cosh^{2}(\gamma_{+}t)}{\Delta_{\sigma}^{2}(t)}>0.
\end{align}
Since $\bm{Y}_{\sigma}(t)$ is a real symmetric
$2\times2$ matrix, its positive trace and determinant establish that
$\bm{Y}_{\sigma}(t)$ is positive definite. Therefore, the regularized
seed remains a square-integrable Gaussian throughout the physical
parameter domain, even when coherent hopping and differential
no-jump backaction produce anisotropy, correlations, and a quadratic phase.

The preceding bounds apply uniformly to the elliptic,
parabolic, and hyperbolic regimes of the resonant dimer. Thus, the
corresponding oscillatory, linear, and exponentially separating reduced
dynamics, respectively, deform the seed without destroying its Gaussian
normalizability anywhere in the physical parameter domain.

\section{Qunaught evolution}
\label{sec:Sec6}

A qunaught is the ideal single-mode GKP grid state with a
one-dimensional code space, with square-lattice phase-space periodicity
and spacing $\sqrt{2\pi}$ \cite{Walshe2020p062411},
\begin{align}
    \langle q \vert \varnothing \rangle \propto&~ \sum_{m\in\mathbb{Z}} \hat{D}\left( \sqrt{2\pi}\,m \right)\delta\left( q \right),
\end{align}
where, for a one-mode configuration displacement,
$\hat{D}(d)=e^{-id\hat{p}}$.
Because its code space is one dimensional, a qunaught carries no logical quantum information by itself.
Its operational role comes from entanglement generation, where interfering two qunaughts on a balanced beam splitter produces the GKP Bell pair used in teleportation-based error correction and computation \cite{Walshe2020p062411}.
Interference of finite-energy qunaughts in two motional modes of a trapped ion
realizes a finite-energy version of this construction \cite{FontboteSchmidt2026}.
The two-mode qunaught product state provides the natural reference state for comparing the lossless beam-splitter construction with its conditioned non-Hermitian deformation.
The corresponding two-mode qunaught product state,
\begin{align}
    \Psi_{\varnothing,\varnothing}\left( q_{1},q_{2},0 \right) \propto&~ \sum_{m,n\in\mathbb{Z}} \hat{D}\left( m\bm{u}_{1} + n\bm{u}_{2} \right)\delta\left( q_{1} \right)\delta\left( q_{2} \right),
\end{align}
with primitive lattice vectors
\begin{align}
    \begin{aligned}
        \bm{u}_{1} =&~ \sqrt{2\pi}\left( 1, 0, 0, 0 \right)^{\mathrm{T}}, \\
        \bm{u}_{2} =&~ \sqrt{2\pi}\left( 0, 1, 0, 0 \right)^{\mathrm{T}},
    \end{aligned}
\end{align}
defines a square Dirac comb in the $\left( q_{1},q_{2} \right)$ plane and admits a Gaussian regularization of its origin seed.

\begin{figure}[tbp]
    \centering
    \includegraphics[width=\linewidth,height=0.53\textheight,keepaspectratio]{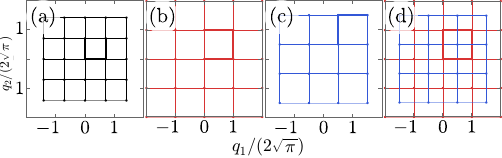}
    \caption{
    Projection of the reference lattices onto the $\left( q_{1},q_{2} \right)$ plane for
    (a) two-mode qunaught product state, (b) the square GKP logical $\lvert 0_{\mathrm{GKP}},0_{\mathrm{GKP}} \rangle$, (c) the square GKP logical $\lvert 1_{\mathrm{GKP}},1_{\mathrm{GKP}} \rangle$, and (d) the GKP$+$ Bell pair $\lvert 0_{\mathrm{GKP}},0_{\mathrm{GKP}} \rangle + \lvert 1_{\mathrm{GKP}},1_{\mathrm{GKP}} \rangle$.
    The axes show the scaled quadratures $q_j/(2\sqrt{\pi})$, and we show finite sections of the infinite grids for visualization.
    }
    \label{fig:Fig1}
\end{figure}

In the lossless limit, $\gamma_{\pm}=0$, our conditioned bosonic dimer reduces to coherent hopping alone and, for $\phi=\pi/2$, the initial primitive vectors evolve as
\begin{align}
    \begin{aligned}
        \bm{u}_{1}\left( t \right) =&~ \sqrt{2\pi}\left( \cos\left( gt \right), \sin\left( gt \right), 0, 0 \right)^{\mathrm{T}}, \\
        \bm{u}_{2}\left( t \right) =&~ \sqrt{2\pi}\left( -\sin\left( gt \right), \cos\left( gt \right), 0, 0 \right)^{\mathrm{T}},
    \end{aligned}
\end{align}
and the evolved regularized seed remains Gaussian.
At the beam-splitter time, $gt=\pi/4$, with primitive vectors
\begin{align}
    \begin{aligned}
        \bm{u}_{1}^{(\mathrm{BS})} =&~ \sqrt{\pi}\left( 1, 1, 0, 0 \right)^{\mathrm{T}}, \\
        \bm{u}_{2}^{(\mathrm{BS})} =&~ \sqrt{\pi}\left( -1, 1, 0, 0 \right)^{\mathrm{T}},
    \end{aligned}
\end{align}
the evolved two-mode qunaught product state becomes
\begin{align}
    \begin{aligned}
        &~\Psi_{\varnothing,\varnothing}^{(\mathrm{BS})}\left( q_{1},q_{2} \right) \\
        &~\propto \sum_{m,n\in\mathbb{Z}} \hat{D}\left( m\bm{u}_{1}^{(\mathrm{BS})} + n\bm{u}_{2}^{(\mathrm{BS})} \right)\delta\left( q_{1} \right)\delta\left( q_{2} \right) \\
        &~\propto \sum_{j,k\in\mathbb{Z}} \hat{D}\left( j\sqrt{2}\bm{u}_{1} + k\sqrt{2}\bm{u}_{2} \right)\delta\left( q_{1} \right)\delta\left( q_{2} \right) \\
        & \quad \, \, + \sum_{j,k\in\mathbb{Z}} \hat{D}\left( \bm{u}_{1}^{(\mathrm{BS})} + j\sqrt{2}\bm{u}_{1} + k\sqrt{2}\bm{u}_{2} \right)\delta\left( q_{1} \right)\delta\left( q_{2} \right) \\
        &~\propto \langle q_{1},q_{2}\vert \bigl( \lvert 0_{\mathrm{GKP}},0_{\mathrm{GKP}} \rangle + \lvert 1_{\mathrm{GKP}},1_{\mathrm{GKP}} \rangle \bigr) .
    \end{aligned}
\end{align}
That is, the two-mode qunaught product state evolves as the superposition of the even-even and odd-odd square GKP sectors, i.e., the GKP$+$ Bell state, where
\begin{align}
    \begin{aligned}
        \langle q \vert 0_{\mathrm{GKP}} \rangle \propto&~ \sum_{j\in\mathbb{Z}} \hat{D}\left( 2\sqrt{\pi}j \right)\delta\left( q \right), \\
        \langle q \vert 1_{\mathrm{GKP}} \rangle \propto&~ \sum_{j\in\mathbb{Z}} \hat{D}\left( \sqrt{\pi}\left( 2j+1 \right) \right)\delta\left( q \right).
    \end{aligned}
\end{align}
This lossless construction provides the reference geometry for the conditioned dynamics, where the same primitive lattice evolves under the full complex symplectic flow and the flow progressively deforms the Bell-pair structure away from the unitary limit.
Figure~\ref{fig:Fig1} displays the configuration-space reference lattices of the qunaught, the square GKP logical states, and the GKP$+$ Bell pair that anchors this reference construction.

\begin{figure}[tbp]
    \centering
    \includegraphics[width=\linewidth,height=0.53\textheight,keepaspectratio]{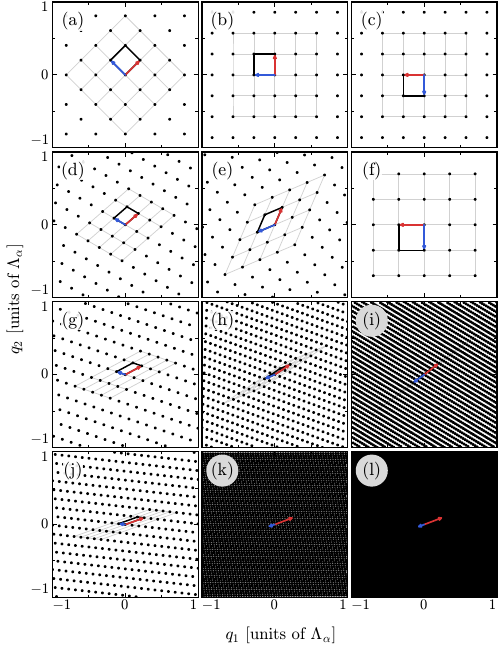}
    \caption{
    Configuration-space projection of the evolved displacement-operator parameters under the conditioned bosonic dimer for (a)--(c) lossless, (d)--(f) elliptic, (g)--(i) parabolic, and (j)--(l) hyperbolic cases.
    Red and blue arrows denote the projected primitive vectors.
    Black dots are all lattice points inside the plotting window, and gray lines mark the $5 \times 5$ block of lattice points around the origin.
    Markers overlap in the densest panels.
    The plotted parameters are the real translation components of each complex displacement; the imaginary momentum components that reweight the displaced seeds are not displayed.
    Axes are scaled by the panel-dependent factors $\Lambda_{\alpha}$ listed in the main text.
    }
    \label{fig:Fig2}
\end{figure}

Figure~\ref{fig:Fig2} shows the configuration-space components of the evolved displacement-operator parameters. Writing
\begin{align}
    \label{eq:DisplacementComponents}
    \bm{S}(t)(m\bm{u}_{1}+n\bm{u}_{2})
    &= \left(\bm{q}_{mn}(t),\bm{p}_{mn}(t)\right)^{\mathrm{T}},
\end{align}
the two-component vectors $\bm{q}_{mn}(t)$ and $\bm{p}_{mn}(t)$ denote, respectively, the configuration- and momentum-displacement components of the evolved displacement vector associated with the lattice point $(m,n)$. For the real configuration and imaginary momentum components considered here, the maximum of each displaced Gaussian component is located at
\begin{align}
    \label{eq:PhysicalPeakPosition}
    \bm{q}_{mn}^{*}(t)
    &= \bm{q}_{mn}(t)-\bm{Y}_{\sigma}^{-1}(t)\operatorname{Im}\bm{p}_{mn}(t).
\end{align}
Thus, despite the apparent expansion of the displacement-parameter lattice in Fig.~\ref{fig:Fig2}, for the dissipative parameter sets shown in Fig.~\ref{fig:Fig2} the physical grid contracts and becomes unresolved once $\gamma_{+}t$ reaches values of approximately $2$--$3$.
Figures~\ref{fig:Fig2}(a)--\ref{fig:Fig2}(c), computed for $\left( \gamma_{1},\gamma_{2} \right)=\left( 0,0 \right)g$, at $t_{\mathrm{BS}}=\pi/(4g)$, $t_{\mathrm{tr}}=\pi/(2g)$, and $t_{\mathrm{rev}}=\pi/g$, give the lossless beam-splitter, transfer, and revival lattices, and make explicit the rotation in the configuration plane characteristic of the bosonic dimer.

Figures~\ref{fig:Fig2}(d)--\ref{fig:Fig2}(f), computed for
$\left( \gamma_{1},\gamma_{2} \right)=\left( 2.4,0.4 \right)g$, at $t_{\mathrm{BS}}=\pi/(4\lvert\kappa\rvert)$, $t_{\mathrm{tr}}=\pi/(2\lvert\kappa\rvert)$, and $t_{\mathrm{rev}}=\pi/\lvert\kappa\rvert$, give the elliptic counterpart,
whose reduced flow is periodic
while the common-loss component of the full complex symplectic flow continues to transform the displacement-parameter lattice. At the reduced revival time, the configuration-space projection therefore retains the factor $\cosh(\gamma_{+}t/2)$.
Figures~\ref{fig:Fig2}(g)--\ref{fig:Fig2}(i), computed for
$\left( \gamma_{1},\gamma_{2} \right)=\left( 4.2,0.2 \right)g$
and evaluated at the same numerical reference times
$t_{\mathrm{BS}}$, $t_{\mathrm{tr}}$, and $t_{\mathrm{rev}}$
defined above for the elliptic case, show the parabolic case at the boundary
between oscillatory and non-oscillatory reduced evolution.
Figures~\ref{fig:Fig2}(j)--\ref{fig:Fig2}(l), computed for
$\left( \gamma_{1},\gamma_{2} \right)=\left( 6.2,0.2 \right)g$
and evaluated at the same numerical reference times
$t_{\mathrm{BS}}$, $t_{\mathrm{tr}}$, and $t_{\mathrm{rev}}$
defined above for the elliptic case, show the hyperbolic case,
with increasing anisotropy of the projected parameter lattice.
The projected area in the $\left( q_{1},q_{2} \right)$ plane changes under the flow, while the four-dimensional transformation remains complex symplectic with unit determinant.
The projected parameter area characterizes these translations, while the physical peaks also depend on the non-unitary weights and seed overlap.
Figure~\ref{fig:Fig2} uses panel-dependent scale factors $\Lambda_{\alpha}$ set by the symmetric plotting windows $q_{1}, q_{2} \in [-\Lambda_{\alpha}, \Lambda_{\alpha}]$, with $\left( \Lambda_{a}, \Lambda_{b}, \Lambda_{c} \right) = \left( 8.74, 8.74, 8.74 \right)$, $\left( \Lambda_{d}, \Lambda_{e}, \Lambda_{f} \right) = \left( 15.87, 23.21, 45.74 \right)$, $\left( \Lambda_{g}, \Lambda_{h}, \Lambda_{i} \right) = \left( 29.07, 120.90, 1567.59 \right)$, and $\left( \Lambda_{j}, \Lambda_{k}, \Lambda_{l} \right) = \left( 65.28, 781.94, 108711.77 \right)$.
The black dots are all lattice points inside each window, and the gray guide lines mark the $5 \times 5$ block of lattice points around the origin.

\begin{figure}[tbp]
    \centering
    \includegraphics[width=\linewidth,height=0.53\textheight,keepaspectratio]{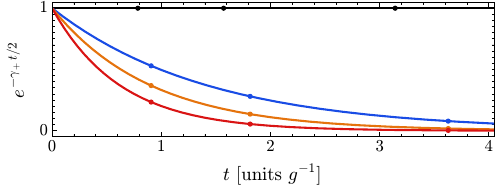}
    \caption{
    Common-mode single-excitation amplitude attenuation factor
    $e^{-\gamma_{+}t/2}$ for the same parameter choices as used in
    Fig.~\ref{fig:Fig2}.
    Shown are the lossless case (black lines and markers), and the elliptic (blue), parabolic (orange), and hyperbolic (red) cases.
    Markers denote the same reference times used in Fig.~\ref{fig:Fig2}.
    This attenuation factor is not the full no-jump
    probability $P_{0}(t)$, which depends on the input state's
    excitation-number distribution and, for $\gamma_{-}\neq0$, on the
    differential-loss dynamics through the reduced propagator. This panel
    isolates only the common-mode amplitude attenuation scale associated
    with one excitation.
    }
    \label{fig:Fig3}
\end{figure}

\begin{figure}[tbp]
    \centering
    \includegraphics[width=\linewidth,height=0.53\textheight,keepaspectratio]{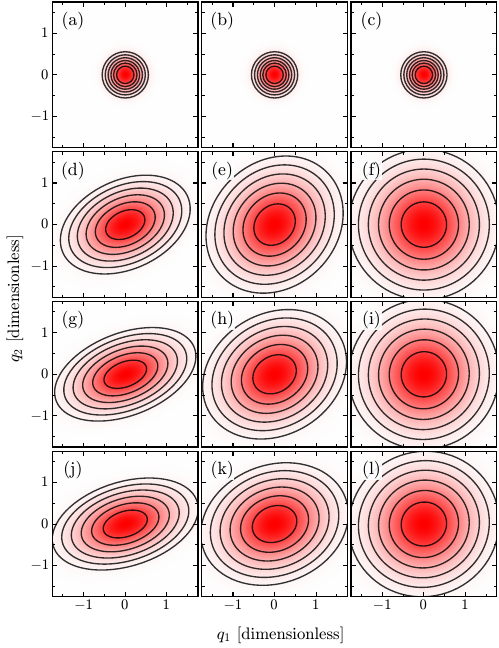}
    \caption{
    Configuration-space density $\lvert \Phi_{\sigma}(\bm{q},t)\rvert^{2}$ of the evolved regularized seed under the conditioned bosonic-dimer dynamics. Each panel is normalized independently to its maximum. For direct comparison with Fig.~\ref{fig:Fig2}, the parameter choices and reference times are kept the same. Specifically, the calculations use resonant modes, $\Delta=0$, hopping phase $\phi=\pi/2$, and initial seed width $\sigma=0.35$. The rows show (a)--(c) the lossless case, $\left(\gamma_{1},\gamma_{2}\right)=\left(0,0\right)g$; (d)--(f) the elliptic case, $\left(\gamma_{1},\gamma_{2}\right)=\left(2.4,0.4\right)g$; (g)--(i) the parabolic case, $\left(\gamma_{1},\gamma_{2}\right)=\left(4.2,0.2\right)g$; and (j)--(l) the hyperbolic case, $\left(\gamma_{1},\gamma_{2}\right)=\left(6.2,0.2\right)g$. From left to right, the lossless panels are evaluated at $t_{\mathrm{BS}}=\pi/(4g)$, $t_{\mathrm{tr}}=\pi/(2g)$, and $t_{\mathrm{rev}}=\pi/g$. The three dissipative rows are evaluated at the common numerical reference times $t_{\mathrm{BS}}=\pi/(4\lvert\kappa_{\mathrm{el}}\rvert)$, $t_{\mathrm{tr}}=\pi/(2\lvert\kappa_{\mathrm{el}}\rvert)$, and $t_{\mathrm{rev}}=\pi/\lvert\kappa_{\mathrm{el}}\rvert$, where $\lvert\kappa_{\mathrm{el}}\rvert=\sqrt{3}g/2$ is the reduced spectral scale for the elliptic parameter set. Red shading represents the normalized density, black lines are its level curves, and the axes show the dimensionless quadratures $q_{1}$ and $q_{2}$.
    }
    \label{fig:Fig4}
\end{figure}

Figure~\ref{fig:Fig3} shows the
common-mode attenuation scale associated with these same parameter choices
and reference times. The plotted factor
$e^{-\gamma_{+}t/2}$ represents the common-mode amplitude attenuation
associated with one excitation; it should not be identified with the full
no-jump probability $P_{0}(t)$, which also depends on the excitation-number
distribution and, for $\gamma_{-}\neq0$, on the differential-loss dynamics.
Figure~\ref{fig:Fig4} shows the corresponding deformation of the
regularized seed for the same parameter choices and reference times.

Figures~\ref{fig:Fig2}--\ref{fig:Fig4} separate three complementary aspects of the conditioned grid evolution. The lattice plots in Fig.~\ref{fig:Fig2} give the
deformation of the displacement parameters that generate the grid
under a determinant-one complex symplectic flow, the attenuation curve in Fig.~\ref{fig:Fig3} gives the common-mode single-excitation amplitude attenuation scale, and the seed plots in Fig.~\ref{fig:Fig4} give the local deformation of the regularized peaks.
In this sense, Fig.~\ref{fig:Fig2} visualizes how the
displacement parameters generating the repeated grid components are
deformed, whereas Fig.~\ref{fig:Fig4} shows how the profile of the
regularized origin seed is modified.
For a normalized finite-energy GKP realization, these effects produce a tradeoff between conditional grid deformation and postselection cost.
The no-jump probability requires a normalized input; the ideal grid alone does not define an absolute postselection probability.
This tradeoff sets the operational constraint for monitored lossy dimers as conditional grid-state transformations.
\section{Experimental platforms}
\label{sec:experiments}

We examine how the model's coupling, loss, and exceptional-point
requirements relate to three leading GKP platforms: superconducting
microwave cavities, trapped-ion motion, and integrated photonics.

Experiments have implemented single-oscillator GKP states in trapped-ion motion, superconducting microwave cavities, and propagating optical fields, while recent trapped-ion experiments have also demonstrated finite-energy GKP gate operations and direct GKP Bell-state generation \cite{Fluhmann2019p513,CampagneIbarcq2020p368,Konno2024p289,Matsos2025p1664}.
Dissipative protocols stabilize finite-energy GKP states in trapped-ion motion \cite{deNeeve2022p296} and superconducting microwave cavities \cite{LachanceQuirion2024p150607}, while a programmable photonic processor generates optical GKP states \cite{Yu2026}.
A technical report describes an experimental two-mode Tesseract grid encoding in a superconducting cavity \cite{NordQuantique2025}.

In superconducting microwave cavities, GKP states occupy storage modes controlled through ancillary circuits \cite{CampagneIbarcq2020p368,Sivak2023p50}.
Related two-cavity devices demonstrate beam-splitter rates $g/2\pi = 319~\mathrm{kHz}$ \cite{Chapman2023p020355} and values exceeding $5~\mathrm{MHz}$ \cite{Lu2023p5767}; the former gives a lossless balanced-beam-splitter time $t_{\mathrm{BS}} = 0.39~\mu\mathrm{s}$.
The cited GKP storage modes have energy lifetimes $T_{1} \in \left[ 245, 606 \right]~\mu\mathrm{s}$, corresponding to loss rates $\gamma = 1/T_{1} \in \left[ 1.65, 4.08 \right] \times 10^{3}~\mathrm{s}^{-1}$ \cite{CampagneIbarcq2020p368,Sivak2023p50}.
The $319~\mathrm{kHz}$ device has its own cavity lifetimes of $482$ and $91~\mu\mathrm{s}$ \cite{Chapman2023p020355}, while the faster converter reports an effective driven single-photon decay time of $197~\mu\mathrm{s}$ at $g/2\pi = 2.16~\mathrm{MHz}$ \cite{Lu2023p5767}.
Using the two lifetimes of the $319~\mathrm{kHz}$ device gives $\lvert \gamma_{-} \rvert/g = 2.2 \times 10^{-3}$ and $\gamma_{+}t_{\mathrm{BS}} = 2.6 \times 10^{-3}$, deep in the elliptic regime of our resonant linear model.
At that coupling, the exceptional-point (parabolic) condition requires
$\lvert \gamma_{1} - \gamma_{2} \rvert = 8.0 \times 10^{6}~\mathrm{s}^{-1}$;
larger differences enter the hyperbolic regime.
If the measured natural losses remained unchanged as the coupling were
reduced, the model's exceptional point would occur at
$g/2\pi \simeq 354.69~\mathrm{Hz}$, where residual nonlinearities and
dephasing must be assessed on the longer interaction time scale.
Superconducting-circuit experiments have also observed postselected
non-Hermitian dynamics by
postselecting trajectories that remain within
an excited-state subspace, including experiments that resolve
exceptional-point signatures in conditioned quantum evolution
\cite{Naghiloo2019p1232,Abbasi2022p160401,Chen2021p140504}.

In trapped-ion motion, finite-energy qunaught interference generates GKP Bell states in two motional modes \cite{FontboteSchmidt2026}.
The beam-splitter coupling reaches $g/2\pi = 1.25~\mathrm{kHz}$, with a $400~\mu\mathrm{s}$ sequence that includes adiabatic ramps.
The grid-interference experiment does not supply a measured pair of pure-loss rates $\gamma_{1,2}$.
Separate reservoir experiments report net cooling rates $\gamma_{c} \in \left[ 1.00, 4.03 \right] \times 10^{3}~\mathrm{s}^{-1}$ and residual heating $\gamma_{h} = 0.43 \times 10^{3}~\mathrm{s}^{-1}$ \cite{So2026p050604}.
Their phonon-loss rate is $\gamma_{c} + \gamma_{h}$ and their phonon-creation rate is $\gamma_{h}$, so the measured net cooling alone does not specify our pure-loss channel.
At the peak beam-splitter coupling, the exceptional-point (parabolic)
condition requires $\lvert \gamma_{1} - \gamma_{2} \rvert =
3.1 \times 10^{4}~\mathrm{s}^{-1}$, with elliptic and hyperbolic
evolution below and above that value.
The demonstrated qunaught interference supplies the coherent reference; approaching or crossing the exceptional point requires engineered differential loss or weaker hopping, with residual heating and dephasing small on the interaction time scale.

Optical experiments generate finite-energy GKP states as propagating modes and in integrated interferometers \cite{Konno2024p289,Larsen2025p587}.
The integrated GKP source reports end-to-end transmission of $78$--$82\%$ from squeezing generation to detection \cite{Larsen2025p587}; this lumped transmission does not determine the loss rates during a subsequent dimer interaction.
Our earlier analysis predicts exceptional points of arbitrary order in a single lossy waveguide beam splitter under photon-number-resolved postselection \cite{QuirozJuarez2019p862}.
A later two-photon waveguide experiment implements a lossy dimer with propagation length replacing time, coupling $g = 0.26~\mathrm{cm}^{-1}$, and a $21~\mathrm{mm}$ coupling section with fan-in and fan-out sections \cite{Klauck2025peadr8275}.
In the waveguide dimer, the reported loss parameter
$\gamma_{\mathrm{wg}}$ ranges from $0$ to $0.63~\mathrm{cm}^{-1}$
and enters the propagation matrix as $-2i\gamma_{\mathrm{wg}}$.
Matching this to the term $-i\gamma_{j}/2$ in our convention gives
a photon-loss difference
$\lvert \gamma_{1} - \gamma_{2} \rvert =
4\gamma_{\mathrm{wg}} \in \left[ 0, 2.52 \right]~\mathrm{cm}^{-1}$.
This range crosses the exceptional point at
$4g=1.04~\mathrm{cm}^{-1}$ and accesses elliptic and hyperbolic
evolution on either side.

An absolute postselection probability cannot be assigned to an ideal grid state because the state is not normalizable.
For a given normalized finite-energy input, the full no-jump probability depends on its excitation-number distribution and, under differential loss, on the reduced dimer dynamics, as discussed in Sec.~\ref{sec:Sec2}.
The analytical bounds derived in Sec.~\ref{sec:Sec2} provide the corresponding postselection range.
A platform-specific numerical probability would also require a finite-energy envelope, interaction protocol, and other platform-dependent parameters.
The experimentally accepted fraction of runs additionally depends on the monitoring scheme and detector response.

\section{Discussion and conclusion}
\label{sec:discussion}

We derived the exact no-jump dynamics of a lossy bosonic dimer with differential decay and used it to study the evolution of ideal grid states.
At the dimer level, the generator separates into a common-mode loss term,
set by the average decay rate, and a reduced non-Hermitian dimer, set by
the interplay of coherent hopping and differential decay.
For resonant modes, the reduced displacement flow is oscillatory in the elliptic regime, linear at the exceptional point, and exponentially separating in the hyperbolic regime.
Detuning generally makes the reduced spectral scale complex and mixes rotation with stretching.
Reaching the exceptional point with passive loss requires
$\lvert \gamma_{1} - \gamma_{2} \rvert = 4g$ and $\gamma_{+} \ge 2g$,
which places the common-mode loss scale on the hopping time scale.

In the geometric sector, the primitive lattice of displacement parameters evolves under a complex symplectic flow in the complexified four-dimensional phase space.
Although the projected area of the lattice changes in the configuration plane, the four-dimensional phase-space transformation remains complex symplectic with unit determinant.
For initial lattice vectors with vanishing momentum components and the beam-splitter hopping phase, the evolved grid combines real configuration translations with non-unitary weights, so its physical peaks depend on the weighted seeds and their interference.
In the seed sector, Gaussian regularization of the origin Dirac-delta
distribution leads to an explicit complex width matrix whose imaginary
part
remains positive definite for any finite, positive
initial width, all non-negative times, and physical loss rates in the
resonant model considered here. Our positivity bounds show that the
regularized seed remains square-integrable and Gaussian.
Its evolution can produce unequal principal widths, correlations,
and a quadratic phase.
For an initial two-mode qunaught product state, the lossless limit recovers the standard beam-splitter generation of a square GKP$+$ Bell pair, while the conditioned no-jump dynamics yields its non-Hermitian deformation.

Measurement-conditioned evolution is an integral part of many quantum-error-correction protocols.
In a monitored bosonic loss channel, postselecting trajectories with no detected loss does not eliminate the conditional backaction associated with the no-jump measurement record.
Our analysis characterizes how this backaction, together with coherent coupling, deforms the displacement-parameter lattice and regularized seed of a grid state.
These effects must be taken into account when assessing postselected protocols or designing recovery operations.
Establishing their consequences for logical performance requires a specified finite-energy encoding, a syndrome-measurement scheme, and a recovery map.
Recovery design and comparisons of logical error rates lie beyond the scope of this work.

Our treatment follows the no-jump branch of ideal grid-state evolution.
Finite-energy grid states under conditional loss are the natural next test of this mechanism.
For an input with known, fixed total excitation number evolving under pure loss,
excitation-number-resolved detection allows postselection on outputs that retain the initial excitation number. This constitutes a distinct postselection strategy from the continuous no-jump trajectory conditioning analyzed above.
Applying this selection to a finite-energy grid requires a protocol that preserves coherence between its excitation-number sectors.
A quantitative treatment requires propagating a specified normalized envelope and accounting for the detector response, heating, dephasing, and residual nonlinearities.
The no-jump probability $P_{0}(t)$ for the normalized input sets the ideal
postselection budget.

\section*{Acknowledgments}
B.~M.~R.~L. thanks Jacinta Alderete Galan for providing daycare support throughout this work.
He also acknowledges support and hospitality as an affiliate visiting colleague at the Department of Physics and Astronomy, University of New Mexico.

\section*{Data availability statement}
All data that support the findings of this study are included within the article (and any supplementary files).

\section*{Funding}
National Science Foundation (PHY-2012172, OSI-2231387); Bayerische Staatsministerium f\"ur Wirtschaft, Landesentwicklung und Energie (6GQT); Bundesministerium f\"ur Bildung und Forschung (16KISK002, 16KIS1598K, 16KISQ039, 16KISQ077, 16KISQ093); Deutsche Forschungsgemeinschaft (1129/2-1).

\section*{Author contributions}
\textbf{B.~M. Rodriguez-Lara} (\href{https://orcid.org/0000-0002-7014-0450}{0000-0002-7014-0450}): Conceptualization (equal), Data curation (equal), Formal analysis (equal), Investigation (equal), Methodology (equal), Project administration (equal), Supervision (equal), Validation (equal), Visualization (equal), Writing -- original draft (equal), Writing -- review \& editing (equal).

\textbf{H. Ghaemi-Dizicheh} (\href{https://orcid.org/0000-0001-9877-2436}{0000-0001-9877-2436}): Investigation (equal), Resources (equal), Validation (equal), Writing -- review \& editing (equal).

\textbf{S. Dehdashti} (\href{https://orcid.org/0000-0001-7713-5462}{0000-0001-7713-5462}): Resources (equal), Visualization (equal), Writing -- review \& editing (equal).

\textbf{A. Hanke} (\href{https://orcid.org/0009-0003-1268-5967}{0009-0003-1268-5967}): Formal analysis (equal), Funding acquisition (equal), Methodology (equal), Resources (equal), Validation (equal), Writing -- review \& editing (equal).

\textbf{A. Touhami} (\href{https://orcid.org/0000-0003-3168-6001}{0000-0003-3168-6001}): Funding acquisition (equal), Writing -- review \& editing (equal).

\textbf{J. N\"otzel} (\href{https://orcid.org/0000-0003-0091-3072}{0000-0003-0091-3072}): Funding acquisition (equal), Writing -- review \& editing (equal).

\end{document}